\documentclass[9pt,twocolumn,twoside]{osajnl}

\journal{ol} 

\setboolean{shortarticle}{true} 

\title{Phase disruption as a new design paradigm for optimizing the nonlinear-optical response.}

\author[1,*]{Rick Lytel}
\author[1]{Sean M. Mossman}
\author[1]{Mark G. Kuzyk}

\affil[1]{Department of Physics and Astronomy, Washington State University, Pullman, Washington  99164-2814}

\affil[*]{Corresponding author: rlytel@wsu.edu}

\dates{Compiled \today}

\ociscodes{(190.4710) Optical nonlinearities in organic materials; (160.4330) Nonlinear optical materials.}

\doi{\url{http://dx.doi.org/10.1364/OL.XX.XXXXXX}}

\begin{abstract}
The intrinsic optical nonlinearities of linear structures, including conjugated chain polymers and nanowires, are shown to be dramatically enhanced by the judicious placement of a charge diverting path sufficiently short to create a large phase disruption in the dominant eigenfunctions along the main path of probability current.  Phase disruption is proposed as a new general principle for the design of molecules, nanowires and any quasi-1D quantum system with large intrinsic response and does not require charge donor-acceptors at the ends.
\end{abstract}

\setboolean{displaycopyright}{true}

\begin{document}

\maketitle
\thispagestyle{fancy}
\ifthenelse{\boolean{shortarticle}}{\abscontent}{}

The design and realization of ultrafast nonlinear optical materials with large responses to optical frequency electric fields remains an active field of pure and applied research\cite{shi15.01,lin13.01,beels12.01,May07.01}.  To date, no general design rules for obtaining large nonlinearities from any structure have been articulated, and the response of materials remains well below that allowed by quantum physics.  In this letter, we propose a general principle that may explain why modern molecules fall short of their potential and use it to demonstrate simple structures with nonlinear responses approaching the physical limits.

The off-resonance electronic nonlinear optical response of a molecule is completely determined by its energy spectrum and transition moments.  Normalized to its maximum value, the sum over states expression for the intrinsic first hyperpolarizability along the x-axis may be written as\cite{kuzyk00.01}
\begin{eqnarray}\label{beta}
\beta_{xxx} &=& 3^{3/4}\left(\frac{\sqrt{m}}{\hbar}\right)^{3}E_{10}^{7/2}N^{-3/2} \left(\frac{|x_{01}|^{2}\bar{x}_{11}}{E_{10}^{2}}+\right. \nonumber \\
&+&\left.\frac{|x_{02}|^{2}\bar{x}_{22}}{E_{20}^{2}}+\left[\frac{x_{01}x_{20}x_{12}}{E_{10}E_{20}}+c.c.\right]+\cdots\right) \\ &\equiv& \beta_{11} + \beta_{22} + [\beta_{12}+c.c] + \dots\nonumber
\end{eqnarray}
where the sum is over all states, $x_{n \ell}$ is the $n, \ell$ element of the position matrix with $\bar{x} = x - x_{00}$, $E_{n0} = E_n - E_0$ is the difference in energy of eigenstates $n$ and $0$, $N$ is the number of electrons and $m$ their mass.  We include as many as 100 energy eigenstates to calculate $\beta$, but systems with large $\beta$ are always well approximated using only three states.  For brevity, we focus on the first hyperpolarizability, though our results also hold for the second hyperpolarizability, $\gamma_{xxxx}$.  For the remainder of this letter, all hyperpolarizabilities are divided by their maximum values and represent intrinsic tensor properties.  It is evident from Eqn. (\ref{beta}) that a system optimized for nonlinear optics will necessarily have eigenfunctions with a large degree of overlap as well as a large change of dipole moment between contributing levels.  This is an essential trade-off to achieve a large response.

It is generally recognized that the hyperpolarizabilities of most molecules fall at least 30 times short of the limits.\cite{kuzyk00.01,kuzyk00.02}. Monte Carlo simulations discovered the optimum energy spectrum for such molecules and provided a strong indicator of the origin of the gap\cite{shafe11.01}.  Optimization of the effective potential energy an electron experiences across the main direction of a quasi-linear molecule showed that the hyperpolarizabilities may be increased by tuning a few parameters\cite{ather12.01}, by modulation of conjugation along a chain\cite{perez09.01}, or by donor-acceptor substitution and insertion of spacers\cite{stefk13.01}.  However, there is no general rule about how to construct the ideal potential energy profile across a molecule to maximize the nonlinear optical response.

In contrast, quantum graph models can sample the vast space of geometries and topologies of quasi-one dimensional structures, leading to structures whose $\beta$ and $\gamma$ approach the fundamental limits\cite{lytel14.01}.  A quantum graph is a collection of one-dimensional line segments, along which electrons flow freely, connected together in a network to constrain particle dynamics to one-dimensional segments within a higher dimensional space.  Quantum graphs are well suited for exploring nonlinear optics because they capture the geometric and topological properties of charge transfer while being exactly solvable systems with quasi-quadratic energy spectra.  Our approach is to identify the essential features of the transition moments and how to tailor them through topological and geometrical properties without the need of a potential on the graph.

We show that quasi-linear structures, such as polymer chains, generate large responses when the component of the ground and first few excited state wavefunctions along the main chain direction have an abrupt \emph{phase disruption} caused by the presence of an alternate pathway for electron flux, such as a short side group, or a local defect such as a Dirac delta potential.  This effect alone generates nonlinearities near the fundamental limits without requiring any additional features such as donor-acceptor groups.  Finite potentials will not lead to a phase disruption, which is quantified by a slope discontinuity of the wavefucntion at one point. The fact that even a small prong, which carries a small fraction of the electron charge, has a profound effect suggests that symmetry breaking in isolation cannot explain the observations.

\begin{figure}\centering
\includegraphics[width=2.4in]{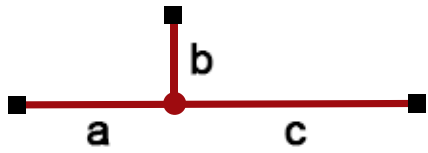}
\caption{The one-prong quantum graph.  All calculations use $a=0.63$ and $c=0.37$. The results are scale-independent.  The horizontal direction is the \emph{main direction} (x axis), and the vertical direction is the \emph{prong direction} (y axis).}\label{fig:fig1}
\end{figure}

\begin{figure}
\center
\includegraphics[width=3.4in]{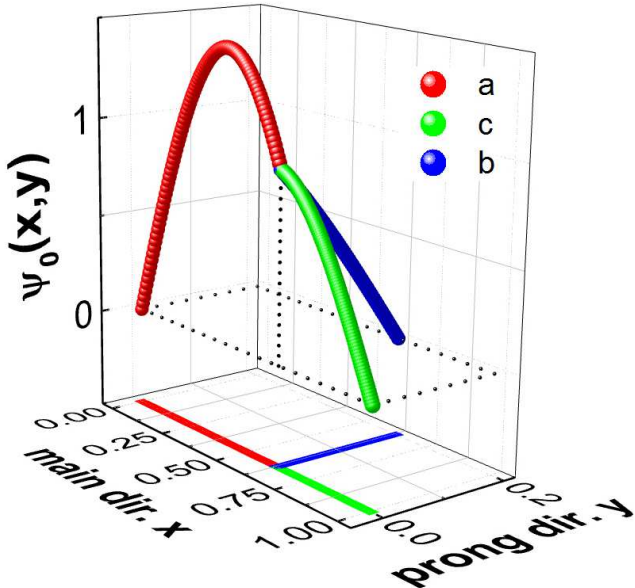}
\vspace{-1em}
\caption{Ground state wavefunction of a one-prong graph.  We label the main direction x and the prong direction y.  The component along the main direction has a kink due to the presence of the prong.}\label{fig:fig2_new}
\vspace{-1em}
\end{figure}

Fig. \ref{fig:fig1} shows a one-prong quantum graph which resembles a linear molecule with a side group normal to the main chain.  The parts of the ground and excited state wavefunctions that lie along the main direction can be tuned to optimum shapes by positioning a short prong to provide a pathway for charge to flow from the main charge-transfer axis, interrupting the flux along the main direction and creating a phase disruption in the wavefunction along that direction.  Fig. \ref{fig:fig2_new} shows the full ground state wavefunction $\psi_{0}(x,y)$ for the one-prong graph, where it is seen that the electron wavefunction along the main direction has a kink caused by the presence of the prong.  We note that computation of quantum graphs requires solutions in both $x$ and $y$ directions\cite{lytel13.01} but the $x$ direction is the dominant contributor to the response for this graph.

\begin{figure}
\includegraphics[width=3.4in]{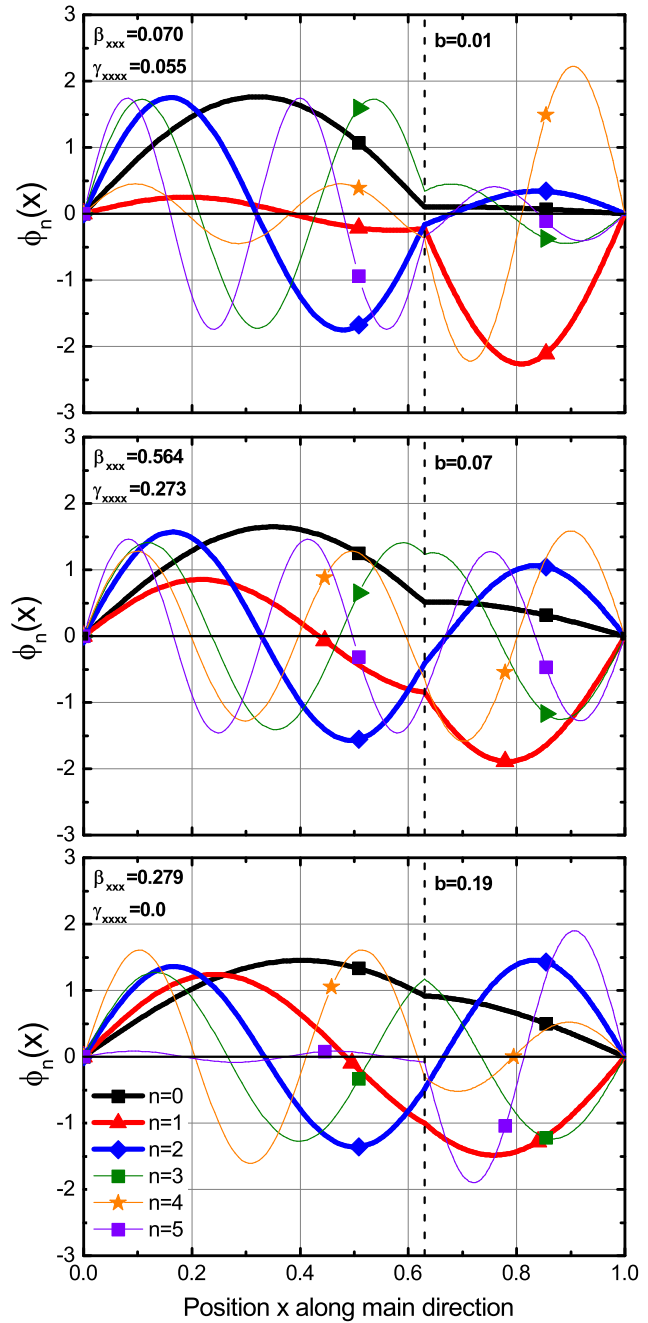}
\caption{The amplitudes $\phi_{n}(x)$ of the wavefunctions along the main direction are plotted for levels $n=0,1,2,3,4,5$ for three prong lengths.  The ground and first two excited states are plotted using thicker lines.  The dashed line indicates the location of prong of length $b$, which is normal to the main direction.  The components of the wavefunctions along the prong are not shown.  The values of $\beta_{xxx}$ and $\gamma_{xxxx}$ are shown for each of the three prong lengths.}\label{fig:fig3_new}
\vspace{-1em}
\end{figure}

Fig. \ref{fig:fig3_new} is the key result and shows how the behavior just described arises.  The part $\phi_{n}(x)=\psi_{n}(x,0)$ of the total wavefunction $\psi_{n}(x,y)$ lies along the main direction, and the first three levels (shown in thicker line type) are the most significant contributors to $\beta_{xxx}$ for this graph topology.  In all three panels, there is a component of the total wavefunction along the prong going into the page at the location of the dashed line.  The length of the prong, $b$, is indicated on each panel.  Also indicated are the values of $\beta_{xxx}$ and $\gamma_{xxxx}$, again normalized to their maximum values.

The top panel of Fig \ref{fig:fig3_new}.  shows the effect of the phase disruption caused by a very short prong.  The wavefunctions are partially localized in such a way that the change in dipole moment, $\bar{x}_{11}$ is quite large when an electron is excited from the ground to the first excited state, but the overlap of the two wavefunctions is nearly zero, making $x_{01}\sim 0$ and eliminating this excitation's contribution to the nonlinearity.  Since this is the largest contributing term to $\beta_{xxx}$, the response is very small, despite the large change in dipole moment:  Both large wavefunction overlap and a large dipole moment change are simultaneous requirements for a significant response.  Similar remarks apply for $\gamma_{xxxx}$, though the falloff with prong length is faster.  Also shown are the next three wavefunctions to illustrate that they, too, have phase disruptions, but their contribution is small due to their much larger excitation energy.

As the prong is lengthened, the middle panel shows that both a large change in dipole moment and good overlap between the first two states persist, indicating large $x_{11}$ and $\bar{x}_{01}$. Consequently, both the first and second hyperpolarizabilities rise significantly:  A short, but not too short, prong results in ideal conditions for large response.  Finally, as the prong is further lengthened, the shapes of the wavefunctions along the main direction approach those of a graph without a prong but with less charge along the main direction due to charge being drawn into the prong.  Thus, the change in dipole moments upon excitation grows smaller, and $\beta_{xxx}$ drops significantly.  But $\gamma_{xxxx}$ is effected more so as it is more sensitive to the drop in changing dipole moments.

\begin{figure}
\includegraphics[width=3.1in]{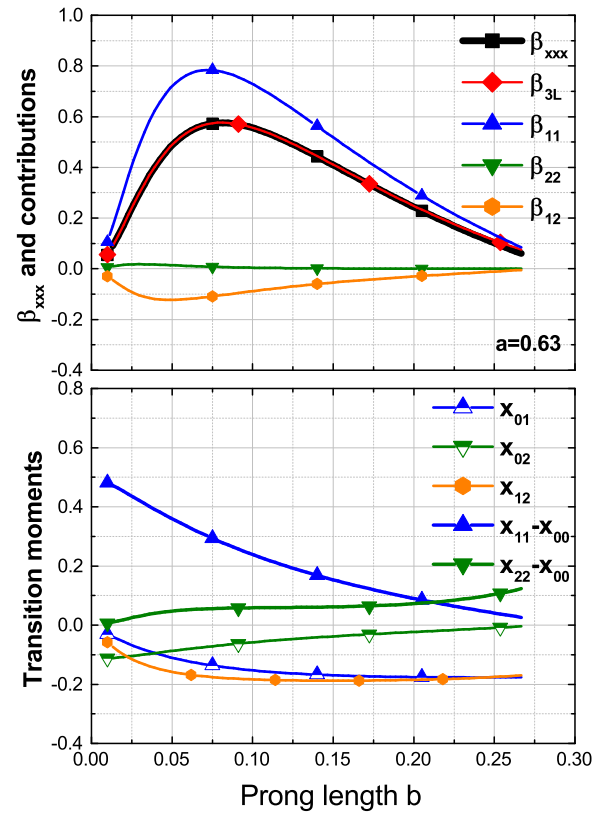}
\caption{The full sum-over-states expression evaluated for $\beta_{xxx}$, the three-level model sum $\beta_{3L}=\beta_{11}+\beta_{22}+\beta_{12}+\beta^{*}_{12}$, and contributions to $\beta_{xxx}$ from the various three-level model terms are shown (top) with corresponding values of the relevant transition moments (bottom) as functions of the prong length. The dependence of the moments on prong length is as expected from the behavior of the wavefunctions functions in Fig. \ref{fig:fig3_new}. (Reprinted with permission of Journal of Nonlinear Optical Physics and Materials.)}\label{fig:fig4_new}
\vspace{-1em}
\end{figure}

We can easily identify the quantitative changes in the relevant transition moments as the prong length is tuned from a short value to a much longer one by examining the first few contributing terms to $\beta_{xxx}$ from Eqn. (\ref{beta}).  Fig. \ref{fig:fig4_new} displays the three lowest-energy contributing terms, their sum, and the full sum over all states for $\beta_{xxx}$ as a function of the prong length $b$ (top), along with the relevant transition and dipole moments as a function of $b$ (bottom).  For very short prongs, the change in dipole moment is largest for the first transition, $\bar{x}_{11}$, but the wavefunction overlap $x_{01}$ of these states is essentially zero.  As the prong length increases, the change in dipole moment drops incrementally but remains large, while the wavefunction overlap moves away from zero, thereby generating a significant contribution to the response. Note that for this graph, the three-level model for $\beta_{xxx}$ is nearly exact.

The analysis reveals the existence of an optimum prong length and position such that the overlap and the change in dipole moment between contributing states are both large.  Short prong lengths of order $7\%$ of the length along the main direction, located approximately $40\%$ from either end, generate a large first hyperpolarizability for the single electron graph, and the response drops dramatically as the prong length is decreased to much shorter values but also as the prong exceeds this optimum value.

\begin{figure}
\includegraphics[width=3.2in]{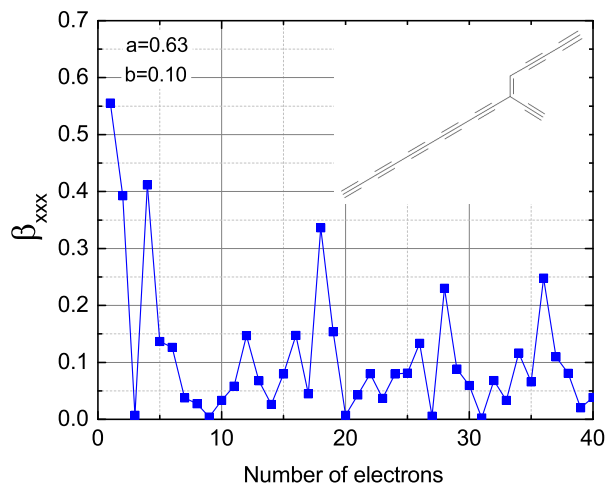} \\
\caption{$\beta_{xxx}$ as a function of electron count $N_e$. The inset shows an 18-electron polyyne molecule whose geometry and topology matches the calculated one, giving $\beta_{xxx} \approx 0.34$.}\label{fig:fig5_new}
\vspace{-1em}
\end{figure}

\begin{figure}
\includegraphics[width=3.2in]{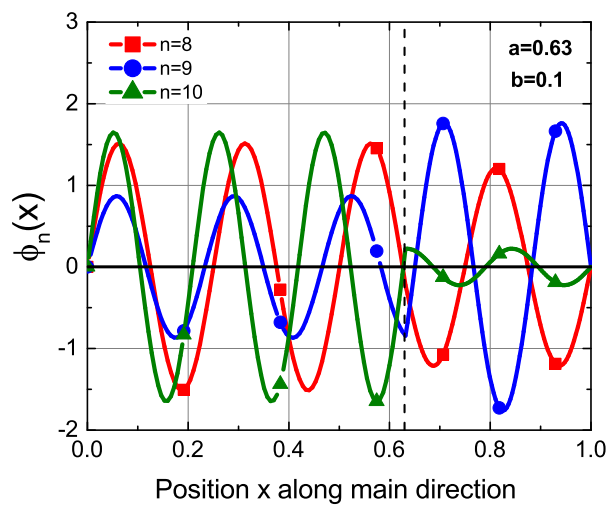}\\
\caption{The amplitudes $\phi_{n}(x)$ of the wavefunctions along the main direction are plotted for single particle states $n=8,9,10$, corresponding to levels at, just below, and just above the Fermi level (state nine) for an 18 electron structure.  Note the large phase disruption in the wavefunctions corresponding to states 9 and 10, leading to large changes in the dipole moment and an enhanced $\beta_{xxx}$, as observed in the previous figure.}\label{fig:fig6_new}
\end{figure}

These results generalize to a non-interacting, many electron system. Fig. \ref{fig:fig5_new} shows $\beta_{xxx}$ as a function of the number of electrons $N_{e}$.  The Fermi level is the highest occupied single electron state.  The large variation in the response with electron count is a consequence of the specific geometry chosen for the graph. {\em For a given number of electrons, there exists a prong position that will maximize the response by optimizing phase disruption of the wavefunctions for states near the Fermi level.} For the graph studied, an electron count of $N_{e}=18$ produces a large response.  Fig. \ref{fig:fig6_new} shows the wavefunctions along the main direction for the states directly below, at, and above the Fermi level with strong phase disruption for wavefunctions corresponding to levels nine and ten. Identical principles to those generating large responses in the one electron quantum graphs apply for many electrons when the electronic states nearest to the Fermi level show phase disruption, as is the case for the molecule in the inset of Fig. \ref{fig:fig5_new}, which we propose as a prototypical large-$\beta$ system.

The general principles delineated here reveal that electronic nonlinear optical molecular designs may be enhanced by judicious placement of a small side group along the main chain of a long molecule.  Large nonlinearities necessarily arise when the most significant terms in the sum-over-states are enhanced by tailoring the shape of the ground and lowest excited state wavefunctions through a phase disruption introduced by this side group when it is positioned closer to the center than to an end of the chain.  Short side groups create significant phase disruption while bleeding off very little charge from the main component of the dipole moment, and at the same time, maintain good wavefunction overlap for side groups that are not vanishingly small.  This behavior strongly supports the view that symmetry breaking alone is not the cause. These results also hold for the second hyperpolarizability suggesting that to any order in nonlinearity promising molecules may be significantly improved by adding a side group off the main chain.

One might argue that quasi-linear structures with infinite potentials at either end and a barrier or well potential in the center might give similar enhancement.  Alternatively, a slant well modeling a donor-acceptor potential across a main chain can also lead to a large response\cite{lytel15.01} without the need of a slope discontinuity in the wavefunctions.  However, it is difficult to engineer a step, bump or slant well potential with a strong-enough potential energy gradient.  In contrast, phase disruption relies only on adding a branch that changes the boundary conditions.

In conclusion, we have shown that a strong phase disruption at a prong's attachment point significantly enhances the nonlinear optical response.  Such structures are simpler to make than using end groups to tilt the potential.  It is likely that molecules with $\beta_{xxx}$ below the gap, designed using somewhat dated principles, do not exhibit the required wavefunction shapes to produce large changes in the dipole matrix elements and, at the same time, maintain large spatial overlap between the contributing eigenstates.  The quantum graph models that we report here, and the corresponding polyyne structure that we propose, should have large phase disruptions corresponding to states near the Fermi level, and thus an ultralarge hyperpolarizability.

\textbf{Funding.} SMM and MGK thank the National Science Foundation (ECCS-1128076) for generously supporting this work.

\bigskip


\end{document}